\documentclass[preprint,aps,prl,showpacs]{revtex4}
\usepackage{graphicx}
\begin{document}


\title{Bond Stiffening in Small Nanoclusters and its Consequences}

\author{Raghani Pushpa}
\author{Umesh Waghmare} 
\author{Shobhana Narasimhan} 
\email{shobhana@jncasr.ac.in}
\affiliation{ Theoretical Sciences Unit,
Jawaharlal Nehru Centre for Advanced Scientific Research,\\ Jakkur PO,
Bangalore 560 064, India}

\begin{abstract}

We have used density functional perturbation theory to investigate the
stiffness of interatomic bonds in small clusters of Si, Sn and Pb.
As the number of atoms in a cluster is
decreased, there is a marked shortening and stiffening of bonds. The
competing factors of fewer but stiffer bonds in clusters result in
softer elastic moduli but higher (average) frequencies as size is
decreased, with clear signatures of universal scaling relationships.
A significant role in understanding trends is played by the
coordination number of the bulk structure: the higher this is, the
lesser is the relative softening of elastic constants, and the greater
the relative damping of vibrational amplitudes, for clusters compared
to the bulk. Our results could provide a framework for understanding recent
reports that some clusters remain solid above the bulk melting temperature.

\end{abstract}

\pacs{63.22.+m, 71.15.Mb, 61.46.Bc}

\maketitle


With the emerging importance of nanotechnology, it has become vital to know 
how the mechanical strength, thermal stability and chemical properties of 
very small objects compare with those of macroscopic size. 
These properties depend crucially on the stiffness of interatomic bonds, 
which determines how difficult it is to move atoms from their equilibrium 
positions - either in thermally induced vibrations, or in response to 
external forces. In this paper, we suggest, using Si, Sn and Pb as examples,
that the shortening and stiffening of bonds in small clusters
may be significant enough to have a noticeable impact on elastic and
thermal properties. We present evidence of some surprising scaling relations,
and also suggest that our results could present a framework for understanding recent
experimental \cite{Shvartsburg, BreauxGa} and computational \cite{Lu, Joshi1, 
Joshi2, Chuang} reports that some
clusters remain solid above the bulk melting temperature, in
contradiction to a long-held belief that small objects will melt at
lower temperatures than the bulk \cite{Schmidt}. 

Low-dimensional systems often display structures where the coordination
number (CN) is less than in the bulk structures of the same element.
From general chemical principles, one expects that such under-coordinated bonds
should be shorter and stiffer; however, the extent of this stiffening is
difficult to estimate accurately from simple arguments. An enhancement in the stiffness of
interatomic bonds has previously been observed in some two- and
one-dimensional systems, e.g., at the surfaces of metals 
\cite{Hoprl,shobfcc110}, in thin
films \cite{Sirenko} and in nanorods and nanotubes \cite{Wong}.
Here, we investigate
trends in bond stiffness, as a function of size and element, in
zero-dimensional clusters that are small enough that their properties
do not obey continuum scaling relations of the bulk and surfaces.
In order to study such effects
theoretically, it is crucial to have a method that can reliably
reproduce the effects of changes in coordination number. Pair
potentials, being insensitive to atomic coordination, are obviously
inadequate, while semi-empirical potentials have to be tailored
carefully if they are to describe such many-body effects not merely
qualitatively but also quantitatively. For these reasons, in this work
we choose to perform quantum mechanical density functional theory (DFT)
and density functional perturbation theory (DFPT) calculations,  using
pseudopotentials and a
plane wave basis, as implemented in the Quantum-ESPRESSO
distribution \cite{pwscf}. Exchange and correlation effects are treated within
the local density approximation (LDA).

\begin{figure}
\includegraphics[width=100mm]{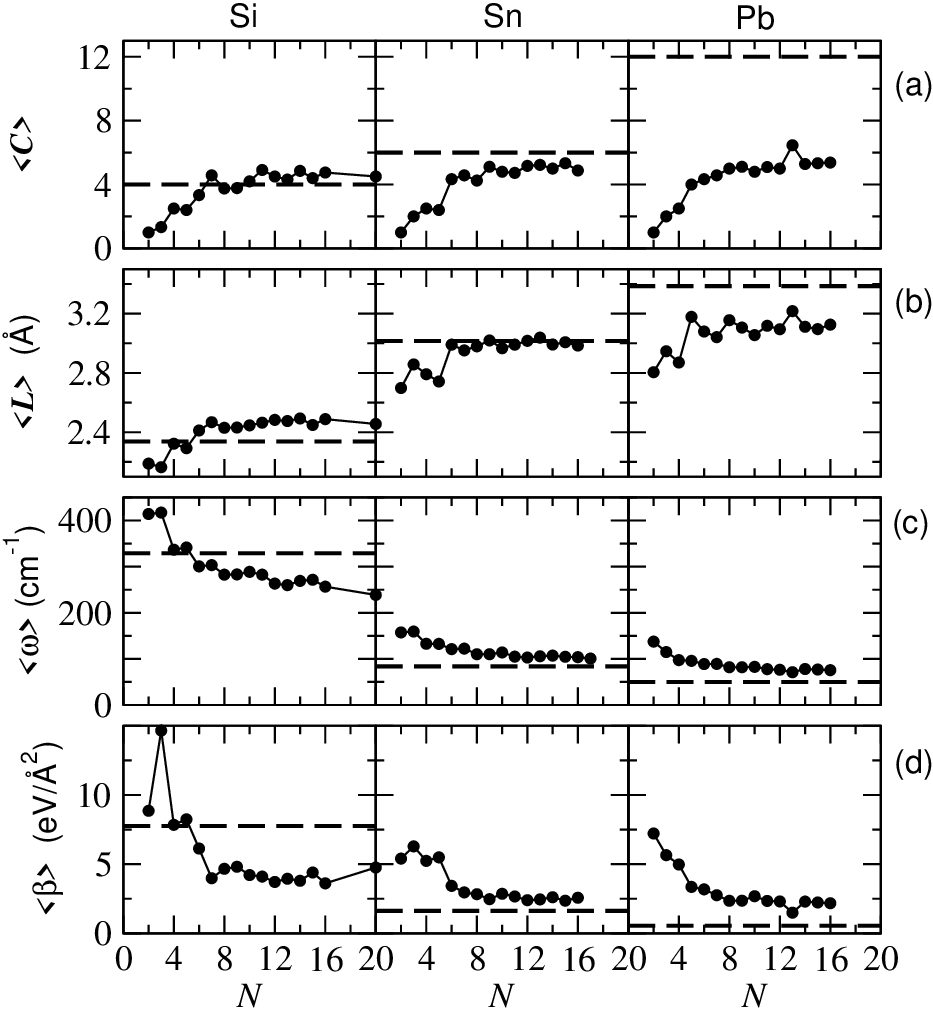}%
\caption{\label{fig1}
Smaller clusters have, on average, lower coordination, shorter and stiffer bonds, 
and higher frequencies. The dots show how the average (a) coordination 
number $\langle C \rangle$, (b) bond length $\langle L \rangle$, (c) 
vibrational frequency $\langle \omega \rangle$, and (d) radial force constant
$\langle \beta \rangle$ vary with $N$, the number 
of atoms in the cluster. The dashed horizontal lines indicate the 
corresponding values for bulk Si, $\beta$-Sn and Pb; note that these lines are positioned 
differently, with respect to the dots, for the three 
elements.}
\end{figure}

We have studied the bulk as well as small clusters (number of atoms $N \leq 20$) of Si,
Sn and Pb. These three elements
belong to the same column of the periodic table, but have different
bulk structures: viz., diamond, {$\beta$}-Sn and face-centred cubic (fcc),
with CNs of 4, 6 and 12 respectively. The clusters we have studied are
tiny enough that they do not resemble bulk fragments structurally.
The
equilibrium structures of clusters were obtained by starting from
previously reported structures \cite{Lu, Ballone, Raghavachari, Honature, 
Bao-xing, Majumder, Doye} and/or regular polyhedral
arrangements, and relaxing using an eigenvector-following technique
\cite{Pushpajcp}
that makes use of eigenvectors obtained using DFPT, and
Hellmann-Feynman forces.
On analyzing the clusters' structures, we find that
both $\langle C \rangle$, the average CN,  
and $\langle L \rangle$, the average bond length, indeed decrease as $N$ becomes smaller
(see Figs.~\ref{fig1}a and \ref{fig1}b).
The shortening is particularly marked for such small clusters,
where all atoms are
essentially surface atoms, and bonds can contract freely, as there 
there is no need to need to maintain registry with a bulk-like core;
this is not true, e.g., for near-surface bonds at
low-index faces of single crystals. It is also noteworthy that the clusters of the three
elements display very similar structures, despite the diverse nature of the
three bulk phases; this may possibly be due to a transition from metallic
to covalent bonding at small cluster sizes, as has been suggested for Ga \cite{Chacko}.

In Fig.~\ref{fig1}a, note
that the dashed line representing the bulk CN is positioned
differently relative to the curve of $\langle C \rangle$ vs. 
$N$ for the three cases: for
Pb, the former lies well above the latter; for Sn, the former lies
slightly above the latter; whereas for Si it is true only for $N\leq 6$
that $\langle C \rangle$ is smaller in the cluster than in the bulk. 
This is of course a
simple corollary of the three different bulk structures of Si, $\beta$-Sn
and Pb; however, its consequences are consistently manifested in three
different kinds of behavior of the clusters relative to the
corresponding bulk, as we will demonstrate below. For example, in Fig.~\ref{fig1}b,
we see that the average bond
length for most Sn (Pb) clusters is less (much less) than the
nearest-neighbor (NN) bond length in the bulk, but for Si this is true
only up to $N=5$.

\begin{figure}
\includegraphics[width=100mm]{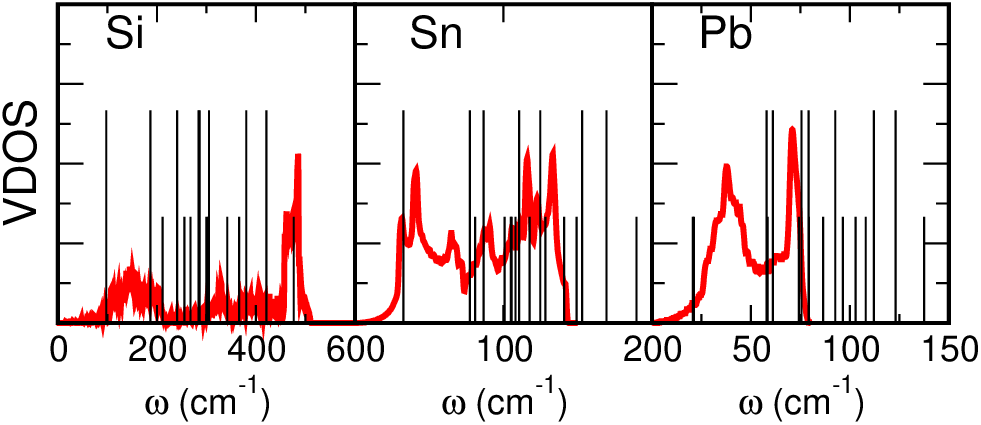}%
\caption{\label{dos_clust_bulk_sisnpb}
DFPT results for the vibrational spectra of clusters and bulk. The thin vertical black 
lines indicate the vibrational density of states (VDOS) for ten-atom clusters 
of Si, Sn and Pb, while the thick curves (red online)
indicate the phonon density of states for 
the corresponding bulk material. The highest 
frequency for the bulk is lower than that of the cluster for Sn and Pb, but not 
for Si.}  
\end{figure}

Next, we 
use DFPT \cite{dfpt} to compute the interatomic force constant tensors (IFCTs),
vibrational frequencies and eigenvectors for all the clusters, as well
as for the bulk structures. We emphasize that this is an exact but
computationally efficient procedure, involving no fitting or
assumptions about the range or form of interatomic interactions, or
about the directions of eigenvectors. Once again, clusters of the
three elements behave differently vis-\`a-vis the bulk: we find that Sn and Pb
clusters have vibrational frequencies that lie above $\omega_{max}^b$, the
highest phonon frequency of the bulk; however, this is not true for Si
clusters (with the exception of $N$=3 and $N$=20). As an example, we
present the vibrational spectrum for ten-atom clusters of Si, Sn and
Pb in Fig.~\ref{dos_clust_bulk_sisnpb}, as well as the corresponding bulk phonon density of
states. Note that the highest frequency mode for Sn$_{10}$ and Pb$_{10}$
exceeds $\omega_{max}^b$ by 32\% and 73\% respectively, whereas for Si$_{10}$ it is
lower by 7\%.

In order to display trends more clearly, we fit the (exact) IFCTs to a sum
of pairwise radial and tangential terms. By assembling the results for all pairs of
atoms for all sizes of clusters,  we have in this way obtained a very large
number of results for radial force constants $\beta$ as a function of bond lengths $L$.
We find that
$\beta$ varies surprisingly smoothly as the inverse eleventh power of $L$, for all
three elements. 
In Figs.~\ref{fig1}c and \ref{fig1}d, we show that the average vibrational frequency 
$\langle \omega \rangle$ and
average radial force constant $\langle \beta \rangle$ increase 
as $N$ is decreased. Note again the three different kinds of behavior relative
to the bulk: in this size range, clusters of Sn and Pb, but
not Si, have stiffer bonds and higher vibrational frequencies (on
average) than the corresponding bulk. It is also clear, on examining the
panels of Fig.~\ref{fig1}, that the size of cluster below
which bonds become stiffer in the cluster than in the bulk is
effectively determined by the coordination number of the bulk
structure.

\begin{figure}
\includegraphics[width=90mm]{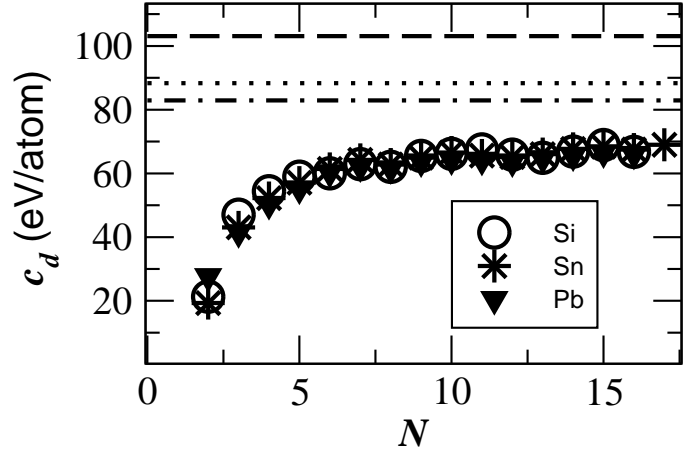}%
\caption{\label{fig3}
Size-dependence of the elastic modulus for dilation, $c_d$. $N$ is  the number of 
atoms in the cluster. The open circles, stars and filled triangles are the data
for Si, Sn and Pb clusters respectively, while the dashed, dotted and dot-dashed
lines represent the results  for bulk Si, $\beta$-Sn and Pb respectively. Note 
that the clusters are softer than the bulk, 
and the data for clusters of the three elements appear to collapse onto a single 
curve.}
\end{figure}

Though the Sn and Pb clusters have stiffer bonds than the
corresponding bulk, there are fewer such bonds per atom. These
two effects compete in determining elastic and thermal properties.
We find that the latter
effect dominates  when we compute the
elastic modulus for dilation, which serves as a measure of
hardness and is defined as   
$c_d = \frac{1}{N}{\partial^2 E}/{\partial \delta^2}$,
where $E$
is the total energy of the system consisting of $N$ atoms, which has
been dilated or compressed by a factor (1+$\delta$); for the clusters, the
dilation was carried out about the centre of mass. 
A striking feature of the graph of $c_d$ vs.~$N$ (Fig.~\ref{fig3}) is
the unexpected data collapse for all but the smallest cluster sizes; we find that
this results
from a scaling relation such that, for a given CN, the bond stiffness multiplied 
by the square of the bond length
is approximately the same for all three elements. Note also that while
all the clusters are softer than the corresponding bulk, Pb clusters
are hardest {\it relative to the bulk} and Si softest, in agreement with the
trends observed above.

\begin{figure}
\includegraphics[width=100mm]{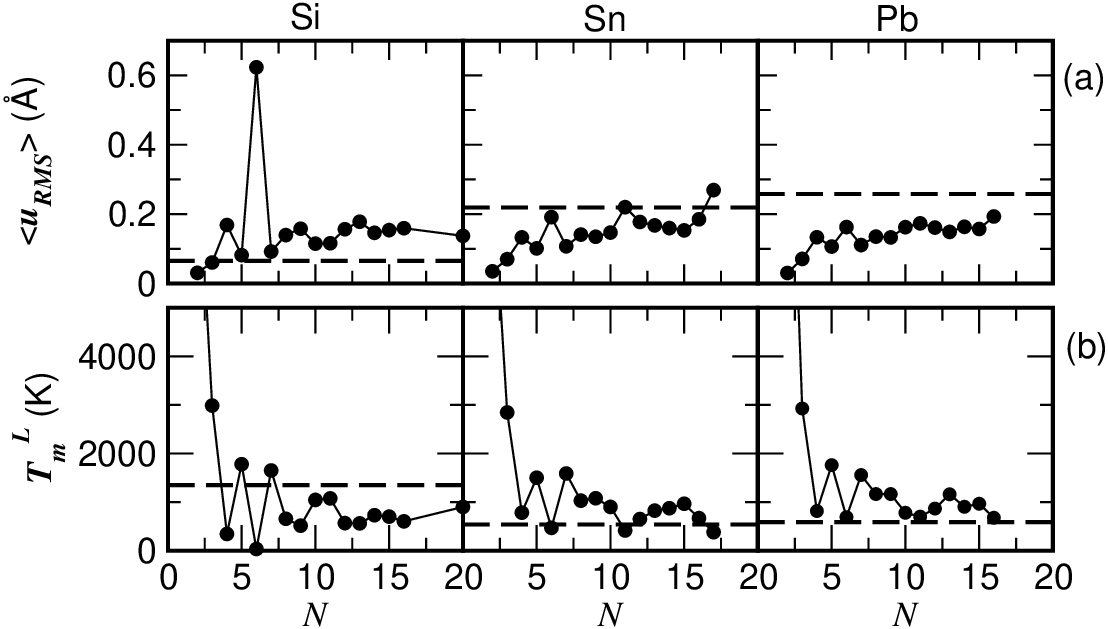}%
\caption{\label{fig4}
Trends and oscillations in vibrational amplitudes and melting temperatures:  
$\langle u_{rms} \rangle$ is the root-mean-squared displacement at 300 K, averaged over all the 
$N$ atoms of the cluster, 
and $T_m^L$ is the Lindemann estimate of the melting temperature.
The dots and dashed lines are the calculated values for the clusters and 
bulk respectively. Note that for Sn and Pb, but not Si, 
the dots lie below (above) the dashed line in the upper (lower) panel.}
\end{figure}

The enhanced bond stiffness competes with the lesser number of bonds
also in determining the mean squared displacements (MSDs) of
atoms; however, in this case, it is the former and not the latter that
wins out.
Within the harmonic approximation, the MSD of the $i^{th}$ atom in
the cluster/bulk, at temperature $T$, is given by \cite{Allen}:

\begin{equation}
\langle u_i^2(T) \rangle = \frac{1}{N_{\bf k}} \sum_{{\bf k}\lambda\alpha}
                               \frac{\hbar}{M\omega_{{\bf k}\lambda}}
                |e^{i\alpha}_{{\bf k}\lambda}|^2 (n_{{\bf k}\lambda} + \frac{1}{2}),
\end{equation}

\noindent
where the vibrational frequencies $\omega_{{\bf k}\lambda}$ and
eigenvectors $e^{i\alpha}_{{\bf k}\lambda}$ are known from DFPT; {\bf k} denotes the phonon wavevector (${\bf k}$=0
for all cluster modes), $N_{\bf k}$ is the number of
wavevectors in the Brillouin zone, $\lambda$ runs over all modes at a given ${\bf k}$,
$\alpha$ specifies Cartesian directions, $M$ is the atomic mass, $\hbar$ is Planck's
constant, and $n_{{\bf k}\lambda}$  is the temperature-dependent Bose-Einstein occupation
factor.
We find that, for a given temperature $T$, (i) though there is
some variation in the MSDs amongst different atoms in a cluster,
there is an overall trend toward smaller MSDs as the cluster size
decreases (see Fig.~\ref{fig4}a); (ii) this variation is however non-monotonic;
(iii) the MSDs for most Sn and Pb (but not Si) clusters are smaller than
for the bulk. This is in contrast to what is observed at the low-index
surfaces of single crystals, where, though the undercoordination of
surface atoms leads to an enhancement in some force constants 
\cite{Hoprl,shobfcc110}, the
MSDs at the surface are still larger than in the bulk
\cite{shobfcc110,shobag110}.

Finally, we investigate the possible implications of our results for
the melting behavior of clusters. The
conventional argument has been that surface atoms have fewer
neighbours than bulk-like atoms, and are therefore less constrained,
resulting in greater amplitudes of thermal vibration, and lower
melting temperatures. This is in accordance with the frequently
observed phenomenon of premelting at flat surfaces of single
crystals \cite{Dosch}, and would suggest that clusters (with their large
surface-to-volume ratio) should melt at lower temperatures than the
bulk. This view was supported by early experiments and
molecular-dynamics simulations on the melting of clusters \cite{Wales}. However,
the majority of these simulations used pair potentials, e.g., Morse or
Lennard-Jones \cite{Wales, Zhou} and thus cannot incorporate any effects of bond
stiffening. Moreover, recent experiments on size-selected clusters of
Sn and Ga suggested that some clusters melt at temperatures above the
bulk melting temperature \cite{Shvartsburg, BreauxGa}. These experiments were initially
motivated \cite{Lu}, and later confirmed \cite{Joshi1,Joshi2,Chuang}, by
{\it ab initio} molecular dynamics
simulations, which showed that certain clusters of Sn and Ga are solid
above the bulk melting temperature $T_m^b$.  However, it has not been
clear whether such results hold only for sizes corresponding to
particularly stable atomic arrangements, or whether they can be
extended to other elements.

As a rough indicator of the consequences of bond stiffening on melting 
behaviour, we have computed the Lindemann melting temperature $T_m^L$, using 
a generalized form \cite{Zhou} of the Lindemann criterion, which states that
objects melt when vibrational amplitudes become equal to a critical
fraction $\Delta$ of interatomic distances:

\begin{equation}
\Delta = \frac{1}{N_b} \sum_{|r_{ij}|<R_c} \frac { \{{\langle u_i^2 \rangle}
             + {\langle u_j^2 \rangle} - {2\langle u_i u_j \rangle} \}^{1/2}}
              {{\langle r_{ij}\rangle}},
\label{delta}
\end{equation}

\noindent
Here, $\langle r_{ij} \rangle$ is the mean value of the distance between 
atoms $i$ and $j$, and $N_b$
is the number of bonds shorter than the cutoff distance $R_c$. We have chosen
$\Delta$ = 0.13 and $R_c$ = 1.15 times the bulk NN distance \footnote{Though the precise
numerical results depend on these choices, we have verified that the trends
displayed are robust.}.
We first computed $T_m^L$ for the bulk phases (dashed lines in Fig.~\ref{fig4}b)
obtaining values of 1350, 537
and 588 K for Si, $\beta$-Sn and Pb respectively, in fairly good agreement
with the experimental melting temperatures of 1680, 505 and 600 K. The
underestimation of $T_m^b$ for Si is a well-known
feature of the LDA, and our estimate is in excellent agreement with
the result obtained using a more sophisticated treatment \cite{Sugino}.

The dots in Fig.~\ref{fig4}b show our results for $T_m^L$ for clusters, as a function of $N$.
(This is only an approximate indicator of melting temperature, 
both because of the empirical nature
of the Lindemann criterion, and because of the broad nature of the melting transition
in finite-sized systems. Moreover, in some clusters, the melting temperature
may be pre-empted by fragmentation \cite{BreauxSn}.) From Fig.~\ref{fig4}b,
we see that huge oscillations are superposed on an overall trend where
$T_m^L$ increases as $N$ decreases. We find that these oscillations in $T_m^L$ result
primarily from oscillations in the value of the lowest vibrational
frequency $\omega_{min}$. Unlike $\langle \omega \rangle$, $\omega_{min}$
 varies
non-monotonically with $N$, is very sensitive to the exact structure,
and reflects variations in tangential force constants. 
These oscillations are reminiscent of those observed
in the size-dependence of the melting temperature of Na clusters \cite{Schmidt},
which could not be explained by either geometric or electronic shell
closing arguments.  

The trends we have observed for the clusters relative to the bulk are maintained 
here too (compare dots and dashed lines in Fig.~4b); i.e.,
for Sn (Pb), most (all) clusters in this size regime have $T_m^L$ above that of the bulk,
whereas for Si, the majority of clusters have $T_m^L$ below the bulk.
For Sn, this is in qualitative agreement with experimental and computational
findings \cite{Shvartsburg, BreauxGa,Lu, Joshi1, 
Joshi2, Chuang}, while we offer our results of enhanced melting temperatures for
Pb clusters as a prediction awaiting experimental validation. 

Interestingly, we find that if the comparison for Sn clusters 
were to be made not with $\beta$-Sn but with the low-temperature
phase of diamond-structure $\alpha$-Sn (which, in reality,
transforms to $\beta$-Sn before it melts), the behaviour of Sn
would be similar to that of Si, i.e., the clusters would have lower vibrational
frequencies and softer bonds, and melt at lower temperatures,
than the bulk.

In summary, we have computed the structure and vibrational properties of
clusters and bulk of Si, Sn and Pb, and shown the presence of clear
size-dependent trends.
The results and analysis presented above suggest persuasively that the
differences (in the comparative behaviour of clusters and bulk)
between Si, Sn and Pb can be attributed to differences in bulk
structure; our arguments are general enough that we believe they
should be valid for a variety of elements. Our results lead to the
following rules-of-thumb: the larger the coordination number in the
bulk, the less the relative softening in the elastic moduli of small
clusters, and the more likely it is that such small clusters are
stable at temperatures above that where the bulk melts. In accordance
with this understanding, we note that a very recent molecular dynamics
simulation \cite{Souledebas} suggests that Au clusters have a melting temperature
above that of bulk (fcc) gold.

\begin{acknowledgments}
Funding was provided by DST.
SN acknowledges the hospitality of the Department of Chemistry, University
of Cambridge, where a part of this work was performed. Helpful conversations with
Dilip Kanhere, David King and Michele Parrinello are gratefully acknowledged.
\end{acknowledgments}
\bibliography{stfclst}


\end{document}